\documentclass[conference]{IEEEtran}
\usepackage{amsmath}
\usepackage{amsfonts}
\usepackage{amssymb}
\usepackage[pdftex]{graphicx}

\newtheorem{theorem}{Theorem}

\newtheorem{definition}[theorem]{Definition}

\renewcommand{\equiv}{\sim}
\newcommand{\vc}[1]{{\mathbf{ #1}}}
\newcommand{\mt}[1]{\mathsf{#1}}
\newcommand{\mat}[1]{\mathsf{#1}}

\renewcommand{\H}{{\mathrm H}}
\newcommand{\D}{{\mathrm D}}

\newcommand{\bR}{{\mathbb {R}}}
\newcommand{\F}{{\mathbb {F}}}

\newcommand{\bZ}{{\mathbb {Z}}}

\newcommand{\DOF}{\mathtt{DOF}}

\newcommand{\ngj}{{\rm NGJV}\ }

\newcommand{\JAP}{\text{JAP}}
\newcommand{\JAPB}{\text{JAP-B}}

\newcommand{\blah}[1]{}

\begin{document}

\title{Delay--Rate Tradeoff in\\ Ergodic Interference Alignment}

\author{\IEEEauthorblockN{Oliver Johnson}
\IEEEauthorblockA{School of Mathematics\\
University of Bristol, UK\\
o.johnson@bristol.ac.uk}
\and
\IEEEauthorblockN{Matthew Aldridge}
\IEEEauthorblockA{Heilbronn Institute of Mathematical Research\\
School of Mathematics\\
University of Bristol, UK\\
m.aldridge@bristol.ac.uk}
\and
\IEEEauthorblockN{Robert Piechocki}
\IEEEauthorblockA{Centre for Communications Research\\
University of Bristol, UK\\
r.j.piechocki@bristol.ac.uk}}
\maketitle

\begin{abstract}
  Ergodic interference alignment, as introduced by Nazer et al (NGJV),
  is a technique that allows
  high-rate communication in $n$-user interference networks with fast fading.
  It works by splitting communication across a pair of
  fading matrices.
  However, it comes with the overhead of a long time delay
  until matchable matrices occur: the delay is $q^{n^2}$ for field
  size $q$.
  
  In this paper, we outline two new families of schemes, called $\JAP$ and
  $\JAPB$, that reduce the expected delay, sometimes at the
  cost of a reduction in rate from the NGJV scheme.
  In particular, we give examples of good schemes for networks with
  few users, and show that in large $n$-user  networks, the delay scales
  like $q^T$, where $T$ is quadratic in $n$ for a constant per-user rate and $T$ is constant for
  a constant sum-rate.  We also show that half the single-user
  rate can be achieved while reducing NGJV's delay from $q^{n^2}$ to
  $q^{(n-1)(n-2)}$.
  
  This extended version includes complete proofs and more details
  of good schemes for small $n$.
\end{abstract}


\section{Introduction}

Interference alignment \cite{cadambe,maddahali} describes a set of techniques
that allow communication in multiuser networks 
at much higher rates than
standard resource division schemes such as TDMA.  Interference alignment schemes
work by `aligning' all interfering signals so they can
be cancelled together.

Given a fast-fading channel -- that is a channel with independent
and identically distributed (IID) fading coefficients in each time
slot -- Nazer, Gastpar, Jafar and Vishwanath \cite{nazer} proposed a
scheme, which we call the \ngj scheme. In brief,
the \ngj scheme pairs together communications using fading matrix $\mt{H}$ with
those using fading matrix $\mt{I} - \mt{H}$,  providing 
a situation with no interference between receivers.
(We outline the NGJV scheme in more detail in Section III.)


The \ngj scheme can be regarded 
as optimal for an $n$-user interference network, since half the single-user
rate is achievable for each user, no matter how large $n$ is \cite{nazer,jafar,johnson28,johnsonc7}. 
However, this is achieved
at the cost of a significant delay in communications.

For definiteness,
we  consider a model of communication over a finite field $\F_q$ of size 
$q$. Since the NGJV scheme requires a particular $n \times n$ channel matrix
with entries in $\F_q \setminus \{ 0 \}$ to occur, the expected delay for a given
message is $(q-1)^{n^2}$, which is roughly $q^{n^2}$ for large $q$.
It is clear that even for $n$ and $q$ relatively small,
this is not a practical delay.  For $n=6$ and $q=3$, for example,
the delay is $2^{36} \approx 7 \times 10^{10}$.

There are five questions we would like to try to answer:
  \begin{enumerate}
    \item Can we find a scheme that, like NGJV, achieves half the single-user
      rate, but has a shorter delay?
    \item Can we find schemes that have shorter delays than NGJV, even
      at some cost to the rate achieved?
    \item Specifically, which schemes from Question 2 perform well for
      situations where we have few users ($n$ small)?
    \item Specifically, which schemes from Question 2 perform well for
      situations where we have many users ($n \to \infty$)?
    \item What is the shortest delay possible for any
      scheme achieving a given rate for a given number of users?
  \end{enumerate}
  
In Section \ref{sec:schemes}, we define a new family of schemes, called
JAP (Subsection IV-A), a beamforming extension JAP-B (Subsection IV-B),
and child schemes derived from them (Section IV-D) 
that have lower time delays than the NGJV scheme for
a variety of different rates, answering Question 2.  As a special case,
the $\JAPB(n)$ schemes (Subsection IV-C)
achieve half the single-user rate, like NGJV, while reducing the delay from
$q^{n^2}$ to $q^{(n-1)(n-2)}$, answering Question 1.  In Section
V, we answer Questions 3 and 4, by finding and analysing the JAP-B schemes
that perform the best for small and large $n$; Table 1
and Figure 1 illustrate the best schemes for small $n$, and Theorems
6 and 7 give the asymptotic behaviour.  Question 5 remains an
open problem, although we do give bounds on the delay achievable for the
best schemes listed above.

Koo, Wu and Gill \cite{koo} have attempted to answer Questions 2 and 3.
We briefly outline their work in Section III.

This is a slightly longer version 
of a paper with the same title \cite{us} presented
at the 2011 IEEE Symposium on Information Theory.
This version includes complete proofs, and extends the results in Table 1 and Figure 1.

\section{Model: The finite field channel} \label{sec:model}

Since ergodic interference alignment relies on matrices being exactly aligned,
Nazer et al \cite{nazer} give their main results in the context of
the finite field channel, where there are only finitely-many
possible fading matrices.  They then use a quantisation argument to
apply their results to the Gaussian case.  In order to allow
comparison of our results, we use the same finite field model.

Each transmitter $i$ has an independent message
$\vc w_i \in \F_q^{m_i}$, which it encodes
as a codeword $(x_i[1],\dots,x_i[N])^\top$ of block length $N$, giving
rate $R_i = (\log q^{m_i})/N = \frac{m_i}{N}\log q$.

At time $t$, receiver $j$ sees channel output
  \begin{equation}
    Y_j[t] = \sum_{i=1}^n H_{ji}[t] x_i[t] + Z_j[t],
  \end{equation}
and needs to decode the message $\vc w_j$.  We can rewrite $(1)$
in matrix form as
  $\vc Y[t] = \mt H[t] \vc x[t] + \vc Z[t]$.
(Throughout, matrices are in sans-serif and vectors in bold.)
We call $\mt H[t]$ the \emph{channel matrix} or \emph{fading matrix}.
As in \cite{nazer}, the noise terms $Z_j[t]$ are IID sequences
from a distribution on $\mathbb{F}_q$ that is a mixture
of a uniform distribution and a
point mass at zero:
  \begin{equation*}
    Z_j = \begin{cases}
            0 & \text{ with prob.\ $1-\rho$,} \\
            z\in \{1, \ldots, q-1 \} & \text{ with prob.\ $\rho/(q-1)$,} \end{cases} \end{equation*}
Like \cite{nazer}, we use an `ergodic' model, where
the channel coefficients $H_{ji}[t]$ are drawn IID and uniformly from
the field $\F_q \setminus \{ 0 \}$
and are redrawn for each time slot. 

We assume 
all transmitters and receivers have full causal channel state information 
for all transmitter--receiver pairs.

By a simple mutual information maximisation, it is easy
to show that the capacity of the single-user finite
field channel $Y = Hx + Z$ for a constant $H \neq 0$ is
$\D(Z) := \log q - \H(Z)$,
where $\H(Z)$ is the entropy of $Z$.

Schemes for the finite field interference
channel often allow each user pair to achieve a fixed fraction of the single-user rate.  We
refer to the `pre-$\D(Z)$ term' as the \emph{degrees of freedom}.

\begin{definition}
  Given an achievable symmetric rate point $(R,\dots,R)$, we
  define the \emph{degrees of freedom} to be
  $\DOF = R/\D(Z)$.
\end{definition}

In particular, a single user can achieve $1$ degree of freedom.

\section{Three existing schemes:\\ NGJV, KWG and TDMA}

The \ngj scheme \cite{nazer} is based on an idea
from previous work of Nazer and Gastpar \cite{nazer2} involving
the performance of finite-field multiple-access channels.

Each receiver's problem takes the form of (1).
Rather than reconstructing a single message $\vc w_j$ (or all
messages $\vc w_1, \vc w_2, \dots, \vc w_n$) as receiver $j$
would normally wish to do, it can be shown
\cite[Theorem 1]{nazer2}, \cite[Lemma 3]{nazer} that receiver $j$
can actually recover the `pseudomessage'
$\sum_{i=1}^n H_{ji} \vc w_i$ at rate $\D(Z)$.  This is done
by all transmitters using the same linear code, so for all $i$, $ (x_i[1], \dots, x_i[N])^\top = \mt{G} \vc w_i$,
for an appropriate $m\times N$ generator matrix $\mt G$.



The \ngj scheme works as follows. Each transmitter $i$
sends two signals encoding the same message $\vc w_i$; first
when the channel matrix is $\mt{H}$,
and second when the channel matrix is $\mt H' := \mt{I} - \mt{H}$. In the first time
slot, each receiver $j$
can decode the pseudomessage
$\sum_{i=1}^n H_{ji} \vc w_i$, using the argument above.  The receiver stores
this pseudomessage in its memory. In 
the second time slot it can decode the pseudomessage
$\sum_{i=1}^n H'_{ji} \vc w_i$, which it also stores in its memory.
The receiver then adds together these two estimates of pseudomessages, to
get its own message
\[ \sum_{i=1}^n H_{ji} \vc w_i + \sum_{i=1}^n H'_{ji} \vc w_i
     = \sum_{i=1}^n ( H_{ji} + \delta_{ji} - H_{ji})  \vc w_i = \vc w_j . \]
Since both pseudomessages can be communiated at rate $\D(Z)$, we conclude that each transmitter
can send its message to its receiver at rate $\D(Z)/2$, since two channel uses are 
needed per message.  This corresponds to $\DOF = 1/2$.


We define the expected time delay for the NGJV scheme to be the average number of time slots we must wait
after seeing a channel matrix $\mt H$ until we see the corresponding matrix $\mt I-\mt H$.  The
time delay is geometrically distributed with mean $1/p$, where $p$ is
the probability that the random channel matrix takes the value $\mt I - \mt H$.
Hence  finding the average
time delay is reduced to finding the probability that the desired
matrix appears in the next time slot.  Since a channel matrix has $n^2$ entries,
each of which needs to take the correct one value of $q-1$ possible values, the
average time delay is
  \begin{equation}
    D = \frac{1}{\big(\frac{1}{q-1}\big)^{n^2}} = (q-1)^{n^2} \equiv q^{n^2}.
  \end{equation}
(Here and elsewhere, we write $f(q) \equiv g(q)$ if $f(q)/g(q) \to 1$
as $q \to \infty$.)

Since this expected delay will be quite large even for modest values
of $q$ and $n$, we will concentrate on the \emph{delay exponent}. 

\begin{definition} An interference alignment scheme with expected
delay $D \equiv C q^T$ for some $C$ and $T$ has \emph{delay exponent}
$T$ and \emph{delay constant} $C$.  Formally,
$T := \lim_{q \to\infty} \log D/\log q$.
\end{definition}

Reduction of the delay exponent is the key aim,
with the delay coefficient playing a secondary role.
Since the finite field model is in some sense an a quantisation
of the continuous case, $q$ should be large enough for accurate quantisation.
Alternatively, for fixed quantisation quality, SNR scales like $q^2$,
so the high-SNR region corresponds to large $q$.
When $q$ is large, the delay exponent $T$ dominates the
delay constant $C$ in determining size of the expected delay $D$.

From (2), we see that the NGJV scheme, which achieves $\DOF=1/2$, 
has a delay exponent of $n^2$.

For comparison, time-division multiple access (TDMA), where
each transmitter--receiver pair has sole access to the channel for an
$n$th of the total time, achieves $\DOF = 1/n$ for an expected delay
$D = n = nq^0$, and hence a delay exponent of $T = 0$.  To an extent,
our new schemes can be seen as `interpolating' between the extremes of
NGJV (high rate, high delay) and TDMA (low rate, low delay).

We also mention some new schemes due to Koo, Wu and
Gill \cite{koo} (KWG).  They attempted to answer Questions 2
and 3, by finding schemes with lower delay than NGJV.
The KWG schemes suggest matching
a larger class of matrices than simply $\mt{H}$ and $\mt{I} - \mt{H}$.
By analysing the hitting probability of an associated Markov chain,
they were able to reduce the expected delay, at the cost of a reduction
in rate and hence degrees of freedom. However, their
schemes only affect the delay by a constant multiple, with
the most successful scheme only reducing the delay to $0.64 (q-1)^{n^2} \equiv 0.64 q^{n^2}$
for a sum-rate of $0.79 \D(Z)$. That is, they only reduce the delay constant $C$, leaving
the delay exponent as $T = n^2$.  For modest $q$ and $n$, say $q=3$, $n=6$, again,
we regard this delay as still impractical.
Since the KWG schemes achieve a lower rate than the NGJV scheme for the
same delay exponent, we need only compare our results with the NGJV scheme.

\section{New alignment schemes: JAP and JAP-B} \label{sec:schemes}

\subsection{The basic scheme: JAP} \label{subsec:NGJVesque}

Write $\vc H_j^{\text{int}}$ for the interference vector
  \[ \vc H_j^{\text{int}} := (H_{j1}, \dots, H_{j\,j-1}, H_{j\,j+1}, \dots, H_{jn}) . \]
  
The NGJV scheme has a long delay as after seeing the matrix $\mat H[t_0]$,
we have to wait for a single matrix $\mat H[t_1] = \mat I - \mat H[t_0]$ to appear
to simultaneously complete the linear dependences 
  \[
    H_{jj}[t_0] + H_{jj}[t_0] = 1, \quad \vc H_j^{\text{int}} [t_0] + \vc H_j^{\text{int}} [t_1] = \vc 0
    \quad \text{for all $j$.}
  \]

We can widen the range of acceptable matrices -- and so reduce the expected delay -- by
  \begin{itemize}
    \item  Building the dependences with more than two matrices;
    \item  Forming any linear combination rather than just a sum;
    \item  Allowing transmitters to complete their dependences at different
             times, rather than simultaneously;
    \item  Recovering a multiple of the desired message, rather than the message itself.
  \end{itemize}

In other words, if there exist scalars $\lambda, \lambda_1, \dots, \lambda_K$ such that
  \begin{align}
    \lambda_0 H_{jj}[t_0] + \lambda_1 H_{jj}[t_1] + \cdots + \lambda_K H_{jj}[t_K] &= \lambda \neq 0 \label{3} \\
    \lambda_0 \vc H_j^{\text{int}}[t_0] + \lambda_1 \vc H_j^{\text{int}}[t_1] + \cdots + \lambda_K \vc H_j^{\text{int}}[t_K] &= \vc 0 , \label{4}
  \end{align}
then receiver $j$ can recover its message from $\mt H[t_0$], $\mt H[t_1], \dots,$ $\mt H[t_K]$
by forming the linear combination of pseudomessages
  \[  \lambda_0 \sum_{i=1}^n H_{ji}[t_0] \vc w_i + \cdots + \lambda_K \sum_{i=1}^n H_{ji}[t_K] \vc w_i = \lambda \vc w_j . \]


The time delay for user $j$ is $t_K - t_0$, and they have communcated
at rate $\D(Z)/(K+1)$ for $\DOF = 1/(K+1)$.

We will need to analyse the probability of \eqref{3} and \eqref{4} holding, in order
to analyse the expected time delay of our schemes. Hence, it will be useful
to note the following lemma.

\emph{Lemma $2 \frac12$:}
Conditional on the interference vectors
$\vc H_j^{\text{int}}[t_0], \vc H_j^{\text{int}}[t_1], \dots, \vc H_j^{\text{int}}[t_K]$
being linearly dependent, the probability that \eqref{3} holds
is $1 - O(q^{-1})$, as $q \to \infty$.

\begin{IEEEproof}
  Assume the interference vectors are linearly dependent.
  That is, assume there exist scalars
  $\lambda_1, \lambda_2 \dots, \lambda_K$ such that
    \[ \lambda_0 \vc H_j^{\text{int}}[t_0] + \lambda_1 \vc H_j^{\text{int}}[t_1] + \cdots + \lambda_K \vc H_j^{\text{int}}[t_K] = \vc 0 \]
  where $L > 0$ of the $\lambda_k$ are nonzero.  Then, \eqref{3} also holds
  provided that the corresponding linear combination
    \begin{equation}
      \lambda_0 H_{jj}[t_0] + \lambda_1 H_{jj}[t_1] + \cdots + \lambda_K H_{jj}[t_K] \label{comb}
    \end{equation}
  is nonzero; call the probability that this happens $p$.
  
  If $\lambda_k \neq 0$, then $\lambda_k H_{jj}[t_k] =: V_k$ is uniform on $\mathbb F_q \setminus \{0\}$;
  and if $\lambda_k = 0$, then $\lambda_k H_{jj}[t_k]$ is always $0$.
  So \eqref{comb} is the sum of $L$ random variables $V_k$ IID uniform on $\mathbb F_q \setminus \{0\}$.
  We can write the mass function of each $V_k$ as
  $(1+\rho)U - \rho \delta_0$,  where $U$
  is uniform on $\mathbb F_q$, $\delta_0$ is a point mass on $0$, and
  $\rho = 1/(q-1)$.  Then the mass function $L$ IID copies is
    \[ (1-(-\rho)^L)U + (-\rho)^L\delta_0 . \]
  Hence, the probability that \eqref{comb} is zero
  is
    \[ 1 - p = \big( 1 - (-\rho)^L) \frac{1}{q} + (-\rho)^L = \frac{1}{q} + \frac{1}{q(q-1)^{L-1}} = O(q^{-1}), \]
  as desired.
\end{IEEEproof}

We now define our first new scheme.

Start by fixing $K\leq n$ and a sequence $\vc a = (a_1, a_2, \dots, a_K)$ of length
$K$ and weight $n$, so in the set
  \[ \mathcal A(n,K) := \bigg\{ \vc a \in \bZ_+^K : \sum_{k=1}^K a_k = n \bigg\}, \]
and write $A_k$ for the partial sums
$A_k := a_1 + \cdots + a_k$.
Then we define the scheme $\JAP(\vc a)$
as consisting of the following $K+1$ steps:
  \begin{itemize}
    \item \textbf{Step 0:} Start with a matrix $\mt H[t_0]$.
    \item \textbf{Step 1:} Set $t_1$ to be the first
      time slot that allows the first $a_1$ receivers
      $1, 2, \dots, A_1$ to recover their message from
      $\mt H[t_0], \mt H[t_1]$.
    \item \textbf{Step \emph{k}:} Set $t_k$ to be the first
      time slot that allows the next $a_k$ receivers
      $A_{k-1}+1, A_{k-1}+2, \dots, A_{k}$ to recover their message from
      $\mt H[t_0], \mt H[t_1], \dots, \mt H[t_k]$.
    \item \textbf{Step \emph{K}:} Set $t_K$ to be the first
      time slot that allows the final $a_K$ receivers
      $A_{K-1}+1, A_{K-1}+2, \dots, A_{K}$ to recover their message from
      $\mt H[t_0], \mt H[t_1], \dots, \mt H[t_K]$.
  \end{itemize}
By the end of this process, all $n=A_K$ receivers have recovered their
message.

Since the message was split over $K+1$ time slots, the common rate of communication
is $\D(Z)/(K+1)$, which corresponds to $\DOF = 1/(K+1)$.


We now examine the delay exponent for our new schemes.

  \begin{theorem}
    Consider the $n$-user finite field interference network.
    Fix $K$ and $\vc a \in \mathcal A(n,K)$.  We use the scheme $\JAP(\vc a)$ as outlined
    above.  Then
      \begin{enumerate}
        \item the expected time for the $k$th round to take place
          is $D \equiv q^{T_k(\vc a)}$, where 
          $T_k(\vc a) = a_k (n-k-1)$;
        \item the delay exponent for the whole scheme is
          \[ T(\vc a) := \max_k T_k(\vc a) = \max_k a_k (n-k-1) . \]
     \end{enumerate}
  \end{theorem}
  
\begin{IEEEproof}
  Recall that the expected delay is the reciprocal of the probability
  the desired match can be made.

  Suppose we are about to begin stage $k$ of a scheme $\JAP(\vc a)$.
  By Lemma $2 \frac12$, provided that
  for the next $a_k$ receivers $\vc H_j^{\text{int}}[t_0],\dots, \vc H_j^{\text{int}}[t_k]$
  are linearly dependent -- ensuring that \eqref{4} holds -- then \eqref{3} holds as well with
  probability $1 - O(q^{-1})$. Since we are only interested in the leading order in $q$, we
  may assume that \eqref{3} will hold.
  
  Also by  Lemma $2 \frac12$, the probability that 
  the first $k$ interference vectors are already linearly dependent,
  is only $O(q^{-1})$.  So again, we may assume they are not.
  
  Write $\mathcal S$ for the span of the first $k$ interference
  vectors of one of the desired $a_k$ receivers $j$, 
    \[ \mathcal S := \text{span} \{ \vc H_j^{\text{int}}[t_0],\dots, \vc H_j^{\text{int}}[t_{k-1}] \} . \]
  The idea is that $\mathcal S$ has size roughly $q^k$, whereas the space of all possible
  interfering has size roughly $q^{n-1}$, giving a probability
  $q^{-(n-k-1)}$ of completing a dependence. But we have to be a little more careful, as
  the $H_{ji}$ cannot take the value $0$.
  
  Specifically, since all possible interference vectors in $(\F_q\setminus\{0\})^{n-1}$
  are equally likely, the probability that the next matrix completes
  a linear dependence is indeed
    \[ \frac{| \mathcal S \cap (\F_q\setminus\{0\})^{n-1}|}{|(\F_q\setminus\{0\})^{n-1}|}
         = \frac{q^k s}{(q-1)^{n-1}} , \]
  where $s$ is the proportion of vectors in $\mathcal S$ with no
  zero entries. By counting the possible coefficients in $\mathbb F_q$ used
  in the span, the inclusion--exclusion formula gives us
    \[ s = 1 - (K-1)\frac{1}{q} + O\left(\frac{1}{q^2}\right) = 1 - O(q^{-1}) . \]
    
  Hence, the probability that the interference vectors are linearly dependent is
    \[
      \frac{q^k s}{(q-1)^{n-1}}  = \frac{q^k \big( 1 - O(q^{-1}) \big)}{(q-1)^{n-1}} 
                  \equiv q^{-(n-k-1)} . \]
    
  This must hold for all $a_k$ receivers, which happens with probability
  $(q^{-(n-k-1)})^{a_k} = q^{-a_k(n-k-1)}$, hence the first result.
  
  For the second result, note that, as $q \to \infty$, the delay
  is dominated by the delay for the slowest round.
\end{IEEEproof}

%

\subsection{Improving delay with beamforming: $\JAPB$}

Beamforming slightly improves the performance of $\JAP(\vc a)$ schemes,
combining ideas from the original Cadambe--Jafar interference alignment
\cite{cadambe} with the JAP scheme.

In round $k$ we can guarantee that the interference matches up for receiver
$l := A_{k-1}+1$.
Each transmitter $i$, instead of repeating their message $\vc w_i$, rather encodes
$( H_{li}[t_k] )^{-1} H_{li}[t_0] \vc w_i$.  (Since the coefficient $H_{li}$ cannot be $0$,
the inverse term certainly exists.)  The total received interferences
at receiver $l$ at times $t_0$ and $t_k$ are
both equal to $\sum_{i \neq l} H_{li}[t_0] \vc w_i$, so can be 
cancelled.

We refer to such schemes that take advantage of beamforming
as $\JAPB(\vc a)$ schemes.

\begin{theorem}
The delay exponent of a $\JAPB(\vc a)$ scheme with
parameter sequence $\vc{a}$ is
  \[ T_B(\vc a) := \max_{k} (a_k-1)(n-k-1) . \]
\end{theorem}

\begin{IEEEproof}
At each round, receiver $A_{k-1}+1$ will automatically
recover its message, leaving the JAP scheme to align interference for
the other $a_k -1$ users.  (Independence of the coefficients $H_{ji}$ ensures
that the other users still have the same problem to solve.)
\end{IEEEproof}

In particular, the $\JAPB$ scheme will always outperform the $\JAP$ scheme
with the same sequence $\vc a$.

\subsection{An interesting special case: $\JAPB(n)$}

An interesting special case of the JAP-B schemes is the case when
$K=1$ and $a_1 = n$; we call this scheme $\JAPB(n)$.

In this case, we have $1/(K+1) = 1/2$ degrees of freedom for a rate of $D(Z)/2$.
From Theorem 4, we see that the delay exponent is
  \[ T_B(n) = (a_1-1)(n-1-1) = (n-1)(n-2) . \]

Effectively, the $\JAPB(n)$ scheme works by using beamforming to
automatically cancel user $1$'s interference, then for users
$2,3,\dots,n$ requiring the existence of diagonal matrices $\mt D_0,
\mt D_1$ such that $\mt D_0 \mt H[t_0] + \mt D_1 \mt H[t_1] = \mt I$.

Note that this is the same rate as is achieved by the original NGJV
scheme, but the delay exponent has been reduced from NGJV's
$n^2$ to $(n-1)(n-2) = n^2 - (3n-2)$.  For small $n$ in particular
this is a worthwhile improvement (see Figure 1). For $n=3$ users, 
where experiments have shown the feasibility of interference
alignment \cite{ayach}, the delay exponent is reduced from $n^2 = 9$ to
$(n-1)(n-2) = 2$.

\subsection{Using time-sharing: child schemes}

Another way to generate new alignment schemes is by time-sharing schemes
designed for a smaller number of users.

Call the NGJV, KWG, JAP and JAP-B schemes `parent schemes'.
Given a parent scheme for an $m$-user network, we can modify
the scheme for use in an $n$-user network for $n>m$, giving what we call a
`child scheme'.

We use TDMA to split an $n$-user network into $\binom nm$ subnetworks,
each of which contains a unique collection of just $m < n$ of the users.
Within each of these $m$-user subnetworks, an $m$-user parent scheme is used, while
the other $n-m$ transmitters remain silent.

An $m$-user child scheme has the same delay exponent as the $n$-user parent scheme,
with the rate, and thus the degrees of freedom, reduced by a factor of
$m/n$.  So an $m$-user JAP-B scheme shared between
$n$ users gives $\DOF = m/n(K+1)$.

In particular, a child scheme from parent NGJV schemes has a
lower delay exponent $m^2 < n^2$ than the
main NGJV scheme, reducing the degrees of freedom
from $1/2$ to $m/2n$.  
As even this outperforms the KWG schemes, we regard this as the benchmark
against which to compare our new JAP-B parent and child schemes.

Child schemes derived from the $\JAPB(n)$ parent scheme are particularly effective, and
usually perform better than other $\JAPB$ schemes, as  we discuss
in the next section.

\begin{table*}[tb]
\renewcommand{\arraystretch}{1.05}
\renewcommand{\tabcolsep}{0.4cm}
\caption{Best $\JAPB(\vc a)$ schemes for small values of $n$ and
$K$: delay exponents (above) and optimal $\vc a$ (below). ($* = {}$non-unique)} \label{tab:smallk}
\begin{center}
\begin{tabular}{c|cccccc}
\hline \\ [-1ex]
             & $n=3$ & $n = 4$ & $n= 5$ & $n = 6$ & $n=7$ & $n=8$ \\[0.1cm]
\hline \\ [-1ex]
$K = 1$      & $2$   & $6$     & $12$  & $20$  & $30$ & $42$   \\
$\DOF = 1/2$ & $(3)$ & $(4)$   & $(5)$ & $(6)$ & $(7)$   & $(8)$  \\[0.25cm]
$K = 2$      & $0$     & $2$       & $4$  & $8$ & $12$    & $18$      \\
$\DOF = 1/3$ & TDMA  & $(1,3)$ & $(2,3)$     & $(3,3)$ & $(3,4)$   & $(4,4)$ \\[0.25cm]
$K = 3$      &       & $0$     & $2$ & $4$   & $6$  & $10$ \\
$\DOF = 1/4$ &       & TDMA    & $(1,1,3)^*$ & $(1,2,3)^*$ & $(2,2,3)$ & $(2,3,3)$  \\[0.25cm]
$K = 4$      & & & $0$ & $2$ & $4$  & $6$  \\
$\DOF = 1/6$ & & & TDMA & $(1,1,1,3)^*$ & $(1,1,2,3)^*$  & $(1,2,2,3)^*$ \\[0.25cm]
$K = 5$ & & & & $0$ & $2$  & $4$  \\
$\DOF = 1/6$ & & & & TDMA & $(1,1,1,1,3)^*$   & $(1,1,1,2,3)^*$ \\[0.25cm]
$K = 6$ & & & & & $0$   & $2$ \\
$\DOF = 1/7$ & & & & & TDMA & $(1,1,1,1,1,3)^*$    \\[0.25cm]
$K = 7$ & & & & &    & $0$ \\
$\DOF = 1/8$ & & & & &   & TDMA  \\[0.1cm]
\hline
\end{tabular}
\end{center}
\end{table*}

\section{Best schemes}\label{sec:best}

\subsection{General case}

Given a number of users $n$ and a desired number of degrees of freedom
$\DOF = 1/(K+1)$, we wish to find a scheme with the lowest delay exponent.

For $K=n-1$ or $n$, when $\DOF = 1/n$ or $1/(n+1)$, the best JAP-B schemes have delay
exponent $0$. This is the same
delay exponent as TDMA, which has $\DOF = 1/n$ also.

For $K \leq n-2$ the best parent scheme will be a
JAP-B scheme with parameter sequence $\vc a \in \mathcal A(n,K)$.  We write $T(n,K)$
for this best delay exponent, that is
  \begin{align*}
    T(n,K) &:= \min_{\vc a \in \mathcal A(n,K)} T_B(\vc a) \\
            &= \min_{\vc a \in \mathcal A(n,K)} \ \max_{k \in \{1,2,\dots,K\}} (a_k - 1)(n-k-1) .
  \end{align*}

We can bound $T(n,K)$ as follows.

\begin{theorem} \label{thm:genbounds}
  Fix $n$ and $K \leq n-2$.
  For $T(n,K)$ as defined above, we have
  have the following bounds:
    \[ \frac{n}{K} (n-1) - (2n-K-1) \leq T(n,K) \leq \frac{n}{K}  (n-2)  .\]
  In particular, for fixed $K$, we have $T(n,K) = n^2/K + o(n^2)$.
\end{theorem}

The gap between the bounds grows linearly with $n$.

The following lemma on partial harmonic sums will be useful.

\emph{Lemma 5$\frac12$:} 
  Let $S(n,K)$ be the partial harmonic sum
    \[ S(n,K) := \sum_{k=1}^K \frac{1}{n - k - 1} = \frac{1}{n-2} +  \cdots + \frac{1}{n-K-1} . \]
  Then we have the bounds
    \[ \frac{K}{n-2} \leq S(n,K) \leq \frac{K}{n-K-1} . \]

\begin{IEEEproof}
  There are $K$ terms in the sum. The largest term is $1/(n-K-1)$; the smallest
  term is $1/(n-2)$.
\end{IEEEproof}

We can now prove Theorem \ref{thm:genbounds}.

\begin{IEEEproof}[Proof of Theorem \ref{thm:genbounds}]
  The value of $T(n,K)$ is lower-bounded by the value of the same
  minimisation problem relaxed to allow the $a_k$ to be real.
  That is,
    \begin{align*}
      T(n,K) &= \min_{\vc a \in \bZ_+^K : \sum_k a_k = n} \ \max_{k \in \{1,2,\dots,K\}} (a_k - 1)(n-k-1) \\
          &\geq \min_{\vc a \in \bR_+^K : \sum_k a_k = n} \ \max_{k \in \{1,2,\dots,K\}} (a_k - 1)(n-k-1) .
    \end{align*}
  The relaxed problem is solved by waterfilling, setting $a_k - 1 = c/(n-k-1)$.
  Requiring the weight of $\vc a$ to be $n$ forces
    \[T(n,K) \geq c = \frac{n-K}{S(n,K)} \geq \frac{(n-K)(n-K-1)}{K} , \]
  where we have used Lemma $5\frac12$.  Rearrangement gives the lower bound.
  
  An upper bound is obtained by using the same $c$ and taking
   \[ a_k - 1 = \left\lceil \frac{c}{n-k-1} \right\rceil \leq \frac{c}{n-k-1} + 1 . \]
  This gives
    \begin{align*}
      T(n,K) &\leq c + \max_k (n-k-1) \\
                  &= \frac{n-K}{S(n,K)} + (n-1-1) \\
               &\leq \frac{(n-K)(n-2)}{K} + (n-2) ,
    \end{align*}
  where we have used Lemma $5\frac12$
  Rearrangement gives the upper bound.
  
  The dominant term in the upper and lower bounds is easily
  seen to be $n^2/K$.
\end{IEEEproof}


\subsection{Few users: Small $n$}

For small values of $n$, we can find the best parent JAP-B schemes
by hand.  (The task
is simplified by noting that the optimal $a_k$ will be nonzero
and increasing in $k$.)
Table \ref{tab:smallk} gives the delay exponents of the
best JAP-B schemes for $n = 3, \dots, 8$ and $K\leq n-2$.

\begin{figure*}[p]
	\centering
		\includegraphics[width=0.8\textwidth]{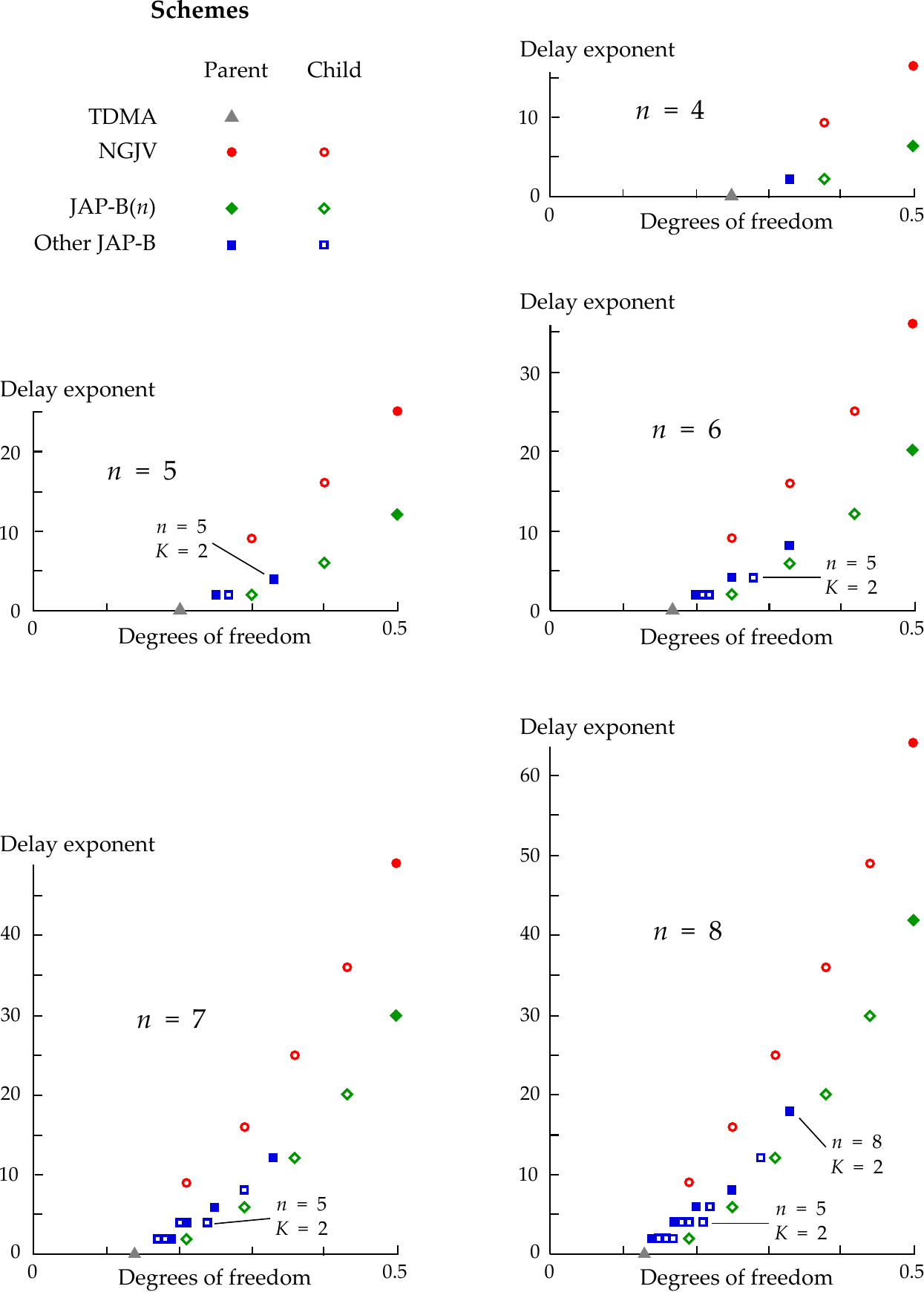}
	\caption{Graphs of delay exponent against degrees of freedom NGJV and best JAP-B parent schemes and child schemes derived from them and TDMA.}
	\label{fig:graphs}
\end{figure*}

We also consider child schemes based on parent JAP-B schemes.
Figure 1 plots the performance of NGJV and all JAP-B schemes, as well
as child schemes derived from them, for $n=3, \dots, 8$.  
For many values of $n$ and $\DOF$, the scheme with the lowest
delay exponent is $\JAPB(n)$ or a child scheme derived
from it, although the JAP-B parent schemes with $(n,K)=(5,2)$ or $(8,2)$, for example, 
and their child schemes always outperform
$\JAP(n)$ for some degrees of freedom.


\subsection{Many users: $n \to \infty$}

We now consider the performance of schemes in the many-user limit
$n \to \infty$.
We consider two limiting regimes, depending on
how the degrees of freedom $\DOF(n)$ should scale with the number of users $n$.
  \begin{itemize}
    \item \textbf{Regime I}, where we hold $\DOF(n) = \alpha$
      constant, for $\alpha \in (0,1/2]$.  That is, we want to communicate at fixed
      fraction of the single-user rate, as in the NGJV scheme.
      NGJV itself corresponds to $\alpha = 1/2$.
  \item \textbf{Regime II}, where we hold the sum-rate
      constant, so the $\DOF(n)= \beta/n$, for $\beta \geq 1$.
      That is, we want to communicate at a multiple of
      the rate allowed by resource division schemes like TDMA.
      TDMA itself corresponds to $\beta =1$.
  \end{itemize}

First, we consider  parent $\JAPB(\vc a)$ schemes.

\begin{theorem}
  Let  $T(n)$ be the delay exponent of the best parent $\JAPB(\vc a)$ scheme for
  $n$-users with at least $\DOF(n)$ degrees of freedom.
  Then:  \begin{itemize}
    \item \textbf{Regime I:} Fix $\alpha\in (0,1/2]$.  Then the delay exponent
      for $\DOF(n) = \alpha$ scales quadratically, in that
        \[ T(n) = \frac{1}{\lfloor 1/\alpha \rfloor - 1} n^2 + o(n^2) = O(n^2) . \]
    \item \textbf{Regime II:} Fix $\beta > 1$.  Then the delay exponent
      for $\DOF(n) = \beta/n$ scales linearly, in that $T(n) = O(n)$, or more specifically
       \[ \left(\beta + \frac{1}{\beta} - 2\right) n - o(n) \leq T(n) \leq \beta n + o(n) . \]
  \end{itemize}
\end{theorem}

\begin{IEEEproof}
  For regime I, we have $1/(K+1) \geq \DOF = \alpha$, so we need to
  take
    \[ K = \left\lfloor \frac{1}{\alpha} - 1 \right\rfloor = \left\lfloor \frac{1}{\alpha} \right\rfloor - 1 . \]
  But 
  Theorem \ref{thm:genbounds} tells
  us that for fixed $K$ we have $T(n,K) = n^2/K + o(n^2)$.
  
  For regime II, we have $1/(K+1) \geq \DOF = \beta/n$, so we need to take
    \[ K = \left\lfloor \frac{n}{\beta} - 1 \right\rfloor = \frac{n}{\beta} - O(1) . \]
  The lower bound from Theorem 5 then gives us
    \begin{align*} T(n) &\geq \frac{n}{n/\beta -O(1)}(n - 1) - \left(2n - \frac{n}{\beta} -O(1)\right) \\
            &=\left (\beta + \frac{1}{\beta} - 2\right) n - O(1), \end{align*}
  and the upper bound gives us
    \[ T(n) \leq \frac{n}{n/\beta - O(1)}(n - 2) = \beta n + O(1).   \]
  This gives the result.
\end{IEEEproof}


In regime I with $\alpha = 1/2$, we get $T(n) \approx n^2$, like NGJV.
  
Figure 1 shows that sharing the parent
scheme $\JAPB(m)$ is particularly effective.  The following theorem
shows this.

\begin{theorem}
  Let $T(n)$ be the delay exponent of child schemes based on $\JAPB(m)$ parent schemes
  for $n$-users with at least $\DOF(n)$ degrees of freedom
  Then:
  \begin{itemize}
    \item \textbf{Regime I:} Fix $\alpha\in (0,1/2]$.  Then the delay exponent
      for $\DOF(n) = \alpha$ scales quadratically, in that
        \[ T(n) =  4 \alpha^2 n^2 + o(n^2). \]
    \item \textbf{Regime II:} Fix $\beta > 1$.  Then the delay exponent
      for $\DOF(n) = \beta/n$ is constant, in that
       \[ T(n) = (\lfloor 2 \beta \rfloor - 1)(\lfloor 2 \beta \rfloor - 2) . \]
  \end{itemize}
\end{theorem}

\begin{IEEEproof}
  Recall from Subsection IV-D that sharing the scheme $\JAPB(m)$ amongst
  $n$ users gives $\DOF = m/2n$ for delay exponent $T = (m-1)(m-2)$.
  
  For regime I, we have $m/2n \geq \DOF(n) = \alpha$, so we need to
  take $m = \lfloor 2 \alpha n \rfloor$, giving
  $T(n) = (\lfloor 2 \alpha n \rfloor - 1)(\lfloor 2 \alpha n \rfloor - 2)$.
  The result follows.
  
  For regime II, we have $m/2n \geq \DOF(n) = \beta/n$, so we need to
  take $m = \lfloor 2 \beta \rfloor$, giving
  $T(n) = (\lfloor 2 \beta \rfloor - 1)(\lfloor 2 \beta \rfloor - 2)$.
\end{IEEEproof}


Note that asymptotically this means that in both regimes child schemes
from $\JAPB(m)$ parent schemes are asymptotically more effective than any other
parent scheme.

Note also that by the same argument as above, sharing the
NGJV parent scheme gives $T(n) = 4\alpha^2 n^2$ in regime I -- less
good than sharing $\JAPB(m)$, but the same to first-order terms.

%
%
%
%
%
%

\section*{Acknowledgments}

M.~Aldridge and R.~Piechocki thank Toshiba Telecommunications Research
Laboratory for supporting this work.
The authors thank Justin Coon and Magnus Sandell for
their advice and support, and Olivier L\'{e}v\^{e}que
for helpful comments.



\end{document}